%% file: main_prd.tex
\documentclass[aps,prd,twocolumn,groupedaddress]{revtex4-1}
\input{preamble}
\usepackage{comment}

\begin{document}

\title{Constraining Super-Heavy Dark Matter with the KM3-230213A Neutrino Event}

\author{Roberto Aloisio~\orcidlink{0000-0003-0161-5923}}
\author{Antonio Ambrosone~\orcidlink{0000-0002-9942-1029}}
\email[]{antonio.ambrosone@gssi.it}
\author{Carmelo Evoli~\orcidlink{0000-0002-6023-5253}}
\affiliation{Gran Sasso Science Institute (GSSI), Viale Francesco Crispi 7, 67100 L’Aquila, Italy}
\affiliation{INFN-Laboratori Nazionali del Gran Sasso (LNGS), via G. Acitelli 22, 67100 Assergi (AQ), Italy}
 
\date{\today}


\begin{abstract}
Recently, the KM3NeT collaboration detected an astrophysical neutrino event, KM3-230213A, with an energy of approximately $220~\rm PeV$, providing unprecedented insights into the ultra-high-energy Universe. In this study, we introduce a novel likelihood framework designed to leverage this event to constrain the properties of super-heavy dark matter (SHDM) decay. Our approach systematically integrates multi-messenger constraints from galactic and extragalactic neutrino flux measurements by IceCube, the absence of comparable neutrino events at IceCube and Auger observatories, and the latest gamma-ray experiment upper limits. Our findings impose the most stringent constraints to date, placing a lower bound on the SHDM lifetime at $\gtrsim 5\cdot 10^{29}-10^{30} \rm s$. Importantly, we identify, for the first time, the significant potential of galactic neutrino flux measurements in advancing dark matter research. Future investigations targeting astrophysical neutrinos originating from the Galactic Center at energies above $10~\rm PeV$ will be crucial, not only for understanding the origin of the cosmic-ray knee but also for exploring the possible contributions of super-heavy dark matter to our Universe.
\end{abstract}


\maketitle

\section{Introduction}
\label{sec:intro}

Dark matter~(DM) is a cornerstone of the standard cosmological model~\cite{Planck:2018vyg}, yet its fundamental nature remains elusive. Over the years, extensive efforts have been devoted to the search for DM particle candidates, spanning a wide mass range from ultra-light ($\sim 10^{-5}\, \rm eV$) to super-heavy ($\sim 10^{21}\, \rm eV$) scenarios~\cite{Bertone:2018krk,Kahlhoefer:2017dnp,Billard:2021uyg,Arguelles:2022nbl,Maity:2021umk,Guepin:2021ljb}. 
In this context, Multi-Messenger Astroparticle Physics provides unique opportunities to probe the properties of DM, as DM annihilation or decay should leave observable signatures in gamma-ray, neutrino, and cosmic-ray data~\cite{Ambrosone:2022mvk,KM3NeT:2024xca,Ajello:2015mfa,IceCube:2024yaw,PierreAuger:2023vql,PierreAuger:2022ubv,PierreAuger:2022jyk}. On the night of February $13^{\mathrm{th}}$, 2023, the KM3NeT/ARCA Cherenkov neutrino telescope in the Mediterranean Sea detected a cosmic neutrino with an energy of approximately $220\,\rm PeV$, designated as the KM3-230213A event~\cite{KM3NeT:2025npi}. This detection may represent a new component in the high-energy neutrino sky, not previously observed by either the IceCube or Pierre Auger Observatories~\cite{IceCube:2025ezc,PierreAuger:2019ens,PierreAuger:2025jwt}. It has been speculated that the KM3NeT-230213A event may provide a hint of physics beyond the Standard Model (BSM)~\cite{Zantedeschi:2024ram,KM3NeT:2025mfl,Barman:2025hoz,Dvali:2025ktz,Jho:2025gaf,Klipfel:2025jql,Telalovic:2025xor,Jiang:2025blz,Alves:2025xul,Turiaci:2025xwi,Narita:2025udw,Kohri:2025bsn,Brdar:2025azm,Boccia:2025hpm,Dvali:2025mqr,Amelino-Camelia:2025lqn,Borah:2025igh,khan2025linkingkm3230213aneutrinoevent,Murase:2025uwv,Bertolez-Martinez:2025trs,Airoldi:2025opo,Anchordoqui:2025xug,Dev:2025czz,Farzan:2025ydi,Baker:2025cff,Khan:2025gxs,Choi:2025hqt,Sakharov:2025oev}, possibly associated with phenomena such as primordial black holes, quantum gravity effects, or the decay of dark matter. In particular, DM candidates with masses above $\sim 10^{8}\,\rm GeV$, collectively referred to as super-heavy dark matter (SHDM), have been proposed as a natural solution to several persisting cosmological problems (see, e.g.,~\cite{Baer:2014eja,Bertone:2004pz} for reviews). However, these interpretations—particularly the SHDM decay scenario—do not account for the fact that the expected signal from SHDM decay should peak toward the Galactic Center, whereas the coordinates of KM3-230213A ($l_{\rm event} = 216.1^\circ$, $b_{\rm event} = -11^\circ$) lie well away from it~\cite{KM3NeT:2025npi}. 
Furthermore, given the energy of the event, cosmogenic neutrinos remain the most viable explanation~\cite{Aloisio:2015ega,KM3NeT:2025vut,AbdulHalim:2023Yd,cermenati}. In this work, assuming that the KM3-230213A event originates from a diffuse astrophysical background, we place limits on the SHDM hypothesis. Unlike previous studies~\cite{Barman:2025hoz,Dvali:2025ktz,Jho:2025gaf,Jiang:2025blz,Narita:2025udw,Kohri:2025bsn,Brdar:2025azm,Borah:2025igh}, we develop a comprehensive likelihood framework that incorporates constraints from ultra-high-energy (UHE) gamma-rays~\cite{PierreAuger:2025jwt,KASCADEGrande:2017vwf,CASA-MIA:1997tns,Fomin:2017ypo,TelescopeArray:2018rbt,Abbasi:2021Z9} as well as neutrino fluxes and upper limits reported by the IceCube collaboration~\cite{IceCube:2023ame,Abbasi:2021qfz}. Moreover, we consistently model both the Galactic contribution, adopting a representative dark matter halo profile, as well as the uniform extragalactic contribution, ensuring a robust treatment of the expected signal across all relevant spatial components. We place the strongest constraints to date on the DM lifetime~$(\gtrsim 5 \times 10^{29}-10^{30}\, \rm s)$. This outcome highlights the critical importance of a global approach for unveiling the properties of a specific dark matter model. Looking ahead, the extension of the galactic neutrino spectrum up to $\sim 10$--$100\,\rm PeV$ will be invaluable in constraining such scenarios, as the Galactic DM-induced contribution is expected to be predominant along the Galactic Plane.

\begin{figure*}[t!]
    \centering
    \includegraphics[width=0.49\linewidth]{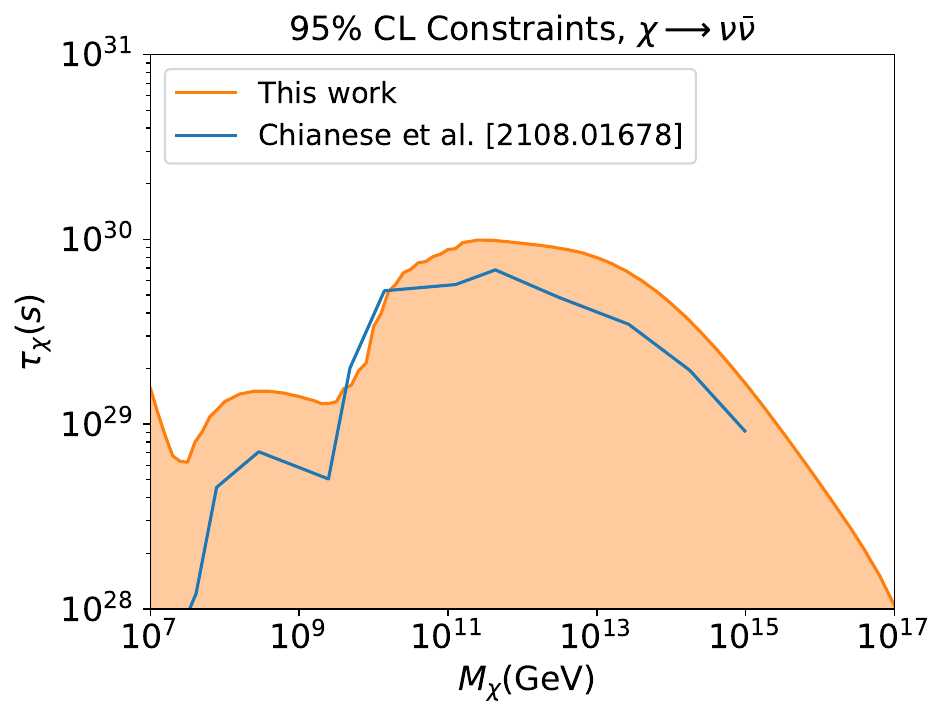}
    \includegraphics[width=0.49\linewidth]{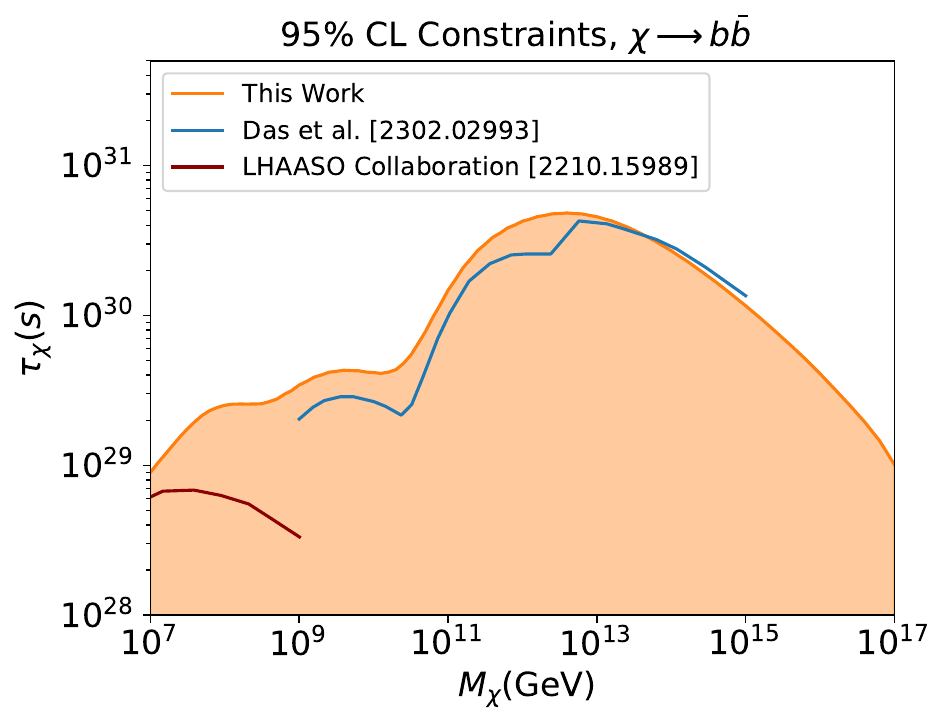}
    \caption{$95\%$ CL lower limit of the SHDM in terms of $M_{\chi}$ in the mass range $10^7-10^{17}\, \rm GeV$. The orange  band represent the excluded regions obtained by our analysis. \textbf{Left}: It refers to $\chi \longrightarrow \nu \bar{\nu}$ and it is compared to the limits obtained by~\cite{Chianese:2021jke}. \textbf{Right}: It refers to $\chi \longrightarrow b \bar{b}$  and it is compared with the limits obtained by~\cite{Das:2023wtk,LHAASO:2022yxw} }
    \label{fig:constraints_DM_lifetime}
\end{figure*}

\section{SHDM Decay Gamma-Ray and Neutrino Fluxes}
\label{sec:fluxes}

The total flux arising from SHDM decay is given by the sum of Galactic and extragalactic contributions~\cite{Chianese:2019kyl}.

The Galactic, per solid-angle flux is given by
\begin{equation}\label{eq:total_flux_galactic}
\begin{split}
\frac{d\phi_{G,i}}{dEd\Omega}\left(E, M_{\chi}, \tau_{\chi}\right) = \frac{1}{4\pi M_{\chi} \tau_{\chi}} \frac{dN_i}{dE}(E) \\ \times \int ds\, \rho(s, l, b)\, e^{-s / \lambda_i(l, b)}~,
\end{split}
\end{equation}
where $i = \gamma, \nu$ denotes either gamma rays or neutrinos, $M_{\chi}$ and $\tau_{\chi}$ are respectively the mass and lifetime of the DM particle~$(\chi)$.  $d\Omega = \cos b\, db\, dl$ is the solid angle element, $\rho(s, l, b)$ is the dark matter energy density (in units of $\rm GeV\,cm^{-3}$), depending on the distance $s$ along the line of sight and Galactic coordinates $(l, b)$. The term ${dN_i (E)}/{dE}$ is the gamma-ray or neutrino energy spectrum per DM decay, computed using the public code \texttt{HDMSpectra}~\cite{Bauer:2020jay}, which accounts for electroweak corrections~\cite{Ciafaloni:2010ti}. $\lambda_i(l, b)$ is the effective attenuation length of gamma rays or neutrinos. Since neutrinos propagate essentially unimpeded up to very high redshift ($z\gtrsim100$), we set $\lambda^{\nu} \rightarrow +\infty$. For gamma rays, the main attenuation is due to pair production on the cosmic microwave background (CMB)~\cite{Das:2023wtk}.
For the Galactic DM energy density, we adopt the Navarro-Frenk-White (NFW) profile~\cite{Navarro:1995iw}, exploiting the parameter values reported in \cite{Chianese:2019kyl}. The Galactic DM density profile remains uncertain and can alternatively be described by other parametrizations, such as the Burkert profile~\cite{Burkert:1995yz,Salucci:2000ps}. However, we have verified that this uncertainty has a negligible effect on our results ($\lesssim 5\%$), and we thus neglect it in the following.

The extragalactic DM distribution also contributes to the overall flux. The differential flux per solid angle from extragalactic decays reads~\cite{Chianese:2019kyl}
\begin{equation}\label{eq:total_flux_extragalactic}
    \frac{d^2 \phi_{EG,i}}{dE\, d\Omega} (E) =  \frac{\Omega_{\chi} \rho_{\rm cr}\, c}{ 4\pi M_{\chi} \tau_{\chi} H_0} \int_{0}^{z_{\rm max}}  \frac{dz}{E(z)}\, \frac{dN_i}{dE}\big(E(1+z)\big),
\end{equation}
where $\rho_{\rm cr} = 4.8 \times 10^{-6}\, \rm GeV\,cm^{-3}$ is the critical density of the Universe, $E(z) = \sqrt{\Omega_M (1+z)^3 + \Omega_\Lambda}$ and $\Omega_{\chi} = 0.26$, $\Omega_M = 0.31$, $\Omega_\Lambda = 0.69$, and $H_0 = 67.74\,\rm km\,s^{-1}\,\rm Mpc^{-1}$ represent the $\Lambda$CDM cosmological parameters as reported by the Planck Collaboration~\cite{Planck:2018vyg}. 
The upper limit of $z_{\rm max} = 5$ for the redshift integral is justified by explicit verification that contributions from higher redshifts are negligible. The gamma ray attenuation severely suppresses the extragalactic contribution to the observable gamma-ray flux. We have verified, using the public \texttt{$\gamma$-Cascade} code~\cite{Capanema:2024nwe}, that secondary cascaded photons do not contribute significantly to the Isotropic Gamma-ray Background (IGRB) measured by Fermi-LAT~\cite{Fermi-LAT:2014ryh}. Consequently, they are not included in the following analysis.
The DM-induced flux is crucially dependent on the source spectrum ${dN}/{dE}$, which in turn depends on the dominant decay channel. In the following, we consider two representative decay channels: i) $\chi \rightarrow \nu \bar{\nu}$ ii) $\chi \rightarrow b \bar{b}$. The first channel corresponds to purely leptonic DM decay, in which the associated gamma-ray signature is highly suppressed at the emission peak~$(\sim M_{\chi}/2)$; arising only from electroweak corrections. By contrast, the second channel represents hadronic DM decay, where the resulting Standard Model particles lead to significant gamma-ray emission through hadronization and subsequent cascade processes. Owing to their qualitatively different phenomenology, these two channels are adopted as representative cases in the following analysis.

\section{On the Direction of the KM3-230213A Event}\label{sec:direction}
\begin{figure*}[t!]
    \centering
    \includegraphics[width=0.6\linewidth]{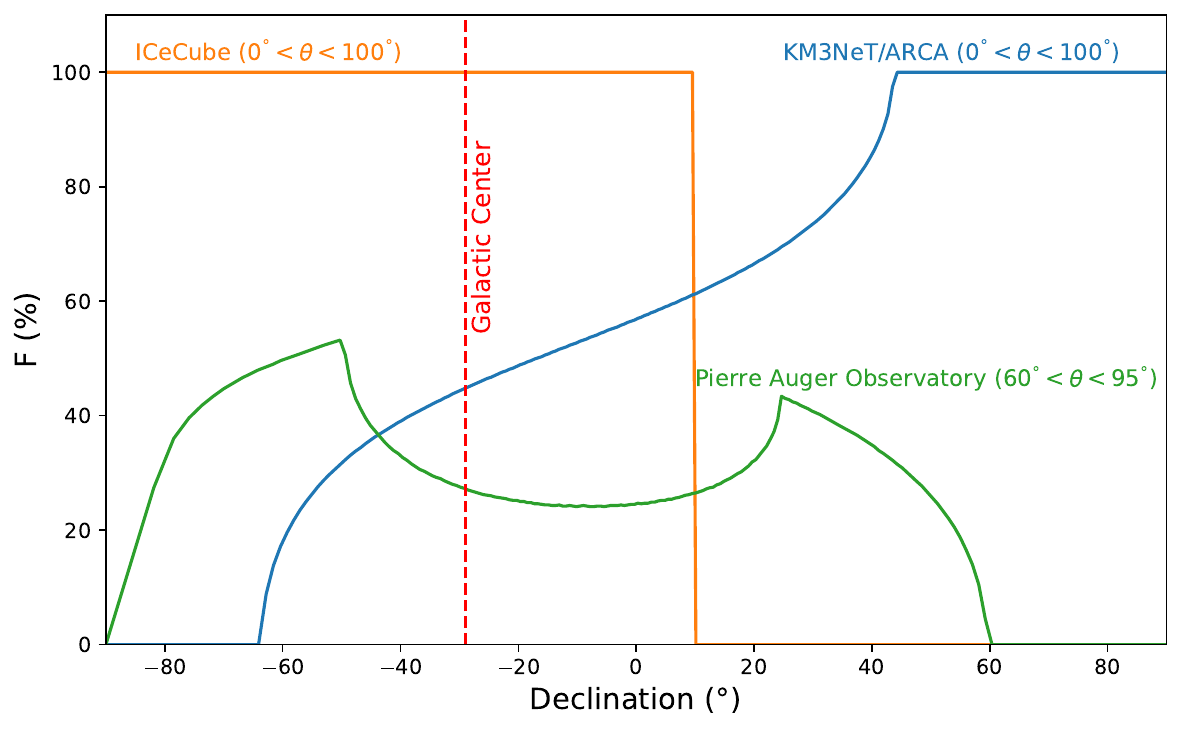}
    \caption{The fractional visibility~$F$ as a function of the equatorial declination for KM3NeT/ARCA~(blue line), IceCube~(orange line) and Pierre Auger Observatory~(green line). The vertical dashed red line refers to the Galactic center declination.  }
    \label{fig:visibility}
\end{figure*}

\begin{figure*}[t]
    \centering
    \includegraphics[width=0.6\linewidth]{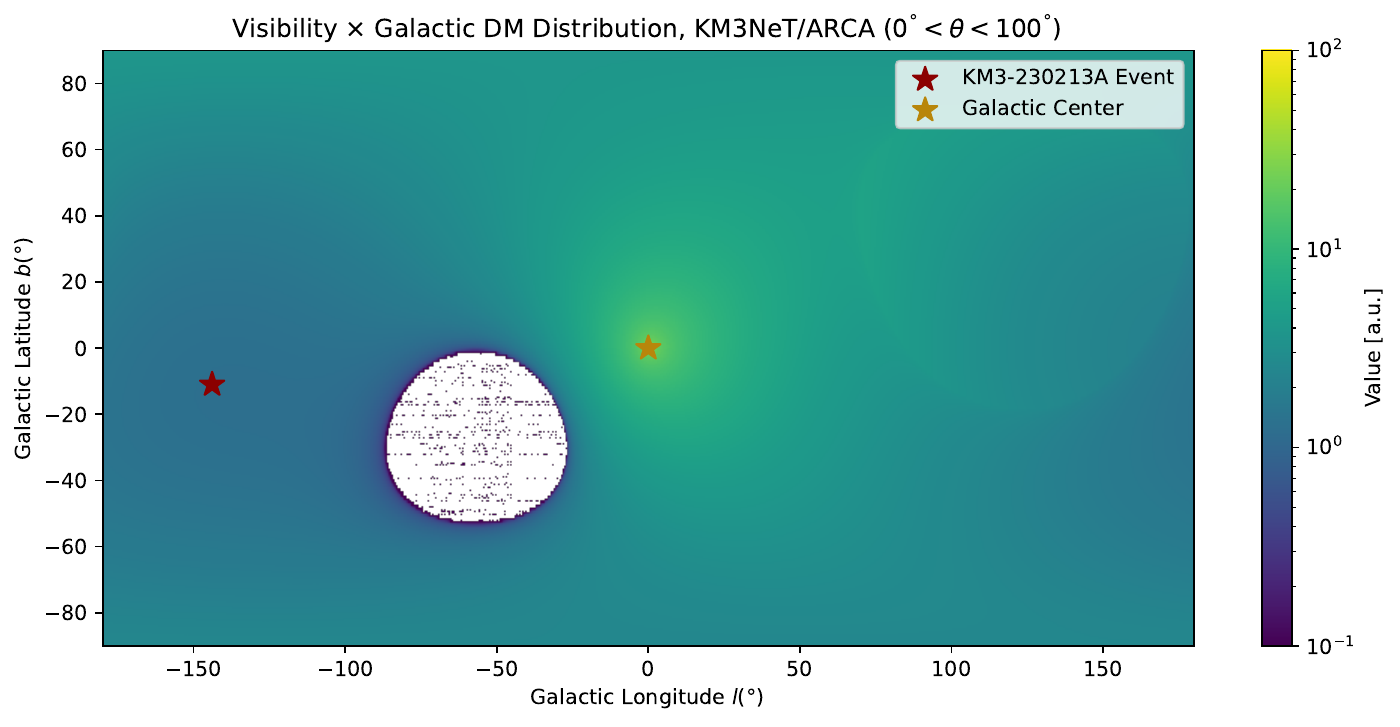}
    \caption{2D map in galactic coordinates of the visibility $\rm F$ $\times$ the galactic DM distribution flux for KM3NeT/ARCA, considering $0^{\circ} \le \theta \le 100^{\circ}$. The blank region corresponds to the blind spots, while the golden and red stars respectively are the Galactic center and the location of KM3-230213A event.}
    \label{fig:final_map_km3}
\end{figure*}

\begin{figure*}[t!]
    \centering
    \includegraphics[width=0.49\linewidth]{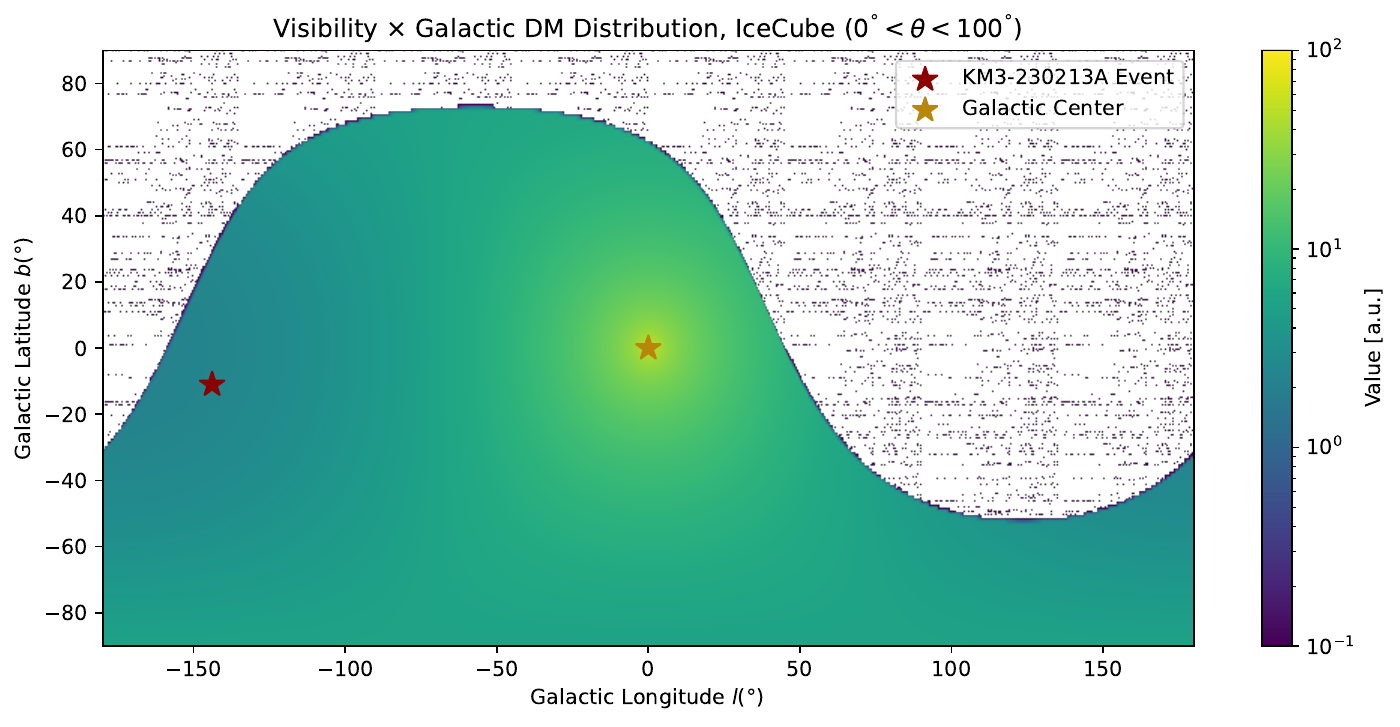}
     \includegraphics[width=0.49\linewidth]{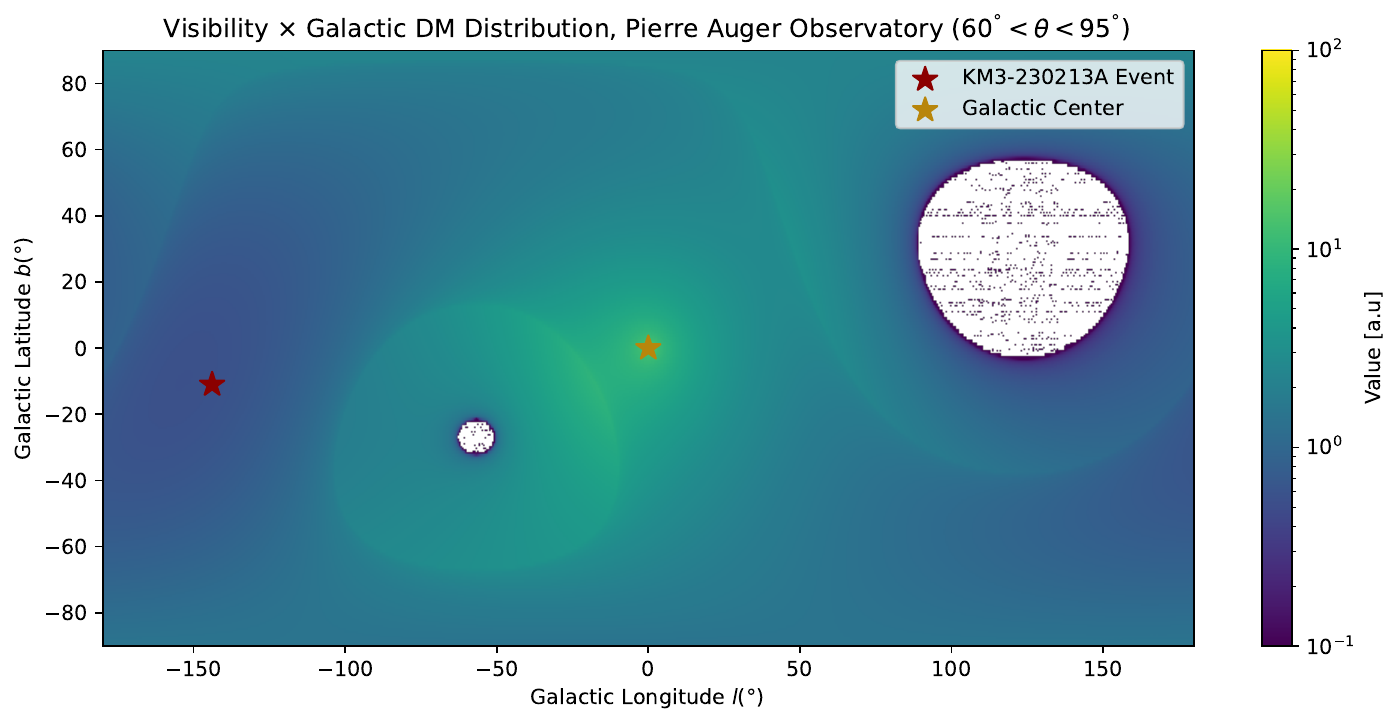}
    \caption{The same as Fig.~\ref{fig:final_map_km3}. \textbf{Left} The IceCube case,  considering $0^{\circ} \le \theta \le 100^{\circ}$. \textbf{Right}: The Pierre Auger case,  considering $60^{\circ} \le \theta \le 95^{\circ}$.}
    \label{fig:final_map_IC}
\end{figure*}

In this section, we demonstrate that the direction of KM3-230213A strongly disfavors a galactic origin, specially in the context of SHDM decay, by exploiting the concept of fraction visibility $F$ of neutrino telescopes.
The visibility of an experiment is defined as the fraction of the day a source, in a given equatorial declination, lies within a given interval of zenith angle~\cite{KM3NeT:2024paj}. This limits the field of view of the experiment and it is simply dependent on the location of the detector.
At ultra high energies~$(E \ge 1-10\, \rm PeV)$, neutrinos can be observed specifically through down-going and horizontal events because  of the Earth absorption~\cite{KM3NeT:2025npi,IceCube:2025ezc}. For KM3NeT/ARCA and IceCube detectors, the corresponding zenith angle range is $0^{\circ} \le \theta \le 100^{\circ}$~\cite{KM3NeT:2024uhg}. For the Pierre Auger Observatory, we consider $60^{\circ} \le \theta \le 95^{\circ}$ according to the range used in \cite{PierreAuger:2019ens}. Fig.~\ref{fig:visibility} reports the fractional visibility as a function of the declination for the experiments, evaluated through the public code available at \url{https://zenodo.org/records/10955399}~\cite{km3net_2024_10955399,KM3NeT:2024paj}. Remarkably, the Galactic Center is $40\%$ of the time observable in the down-going/horizontal sky for KM3NeT/ARCA. Consequently, The Galactic Center should anyway dominate in case of SHDM decay, where the flux is principally concentrated.
To quantify this effect further, Figs.~\ref{fig:final_map_km3} and~\ref{fig:final_map_IC} show 2D maps in galactic coordinates of the product between the visibility and the line-of-sight integral $\int ds, \rho(s, l, b)$, representing the expected contribution from Galactic dark matter decay for KM3NeT/ARCA, IceCube, and the Pierre Auger Observatory, respectively. 


In all cases, the maps are clearly dominated by the Galactic Center. Although detector response and event selection are not included, this trend highlights the central role of the Galactic Center in shaping the expected signal.

\section{Constraints and Analysis Setup}
\label{sec:analysis_setup}


Our analysis evaluates the constraining power of gamma-rays and neutrinos fluxes into constraining the SHDM decay scenario, by enforcing the following constraints:
\begin{itemize}
\item The total neutrino flux (Galactic plus extragalactic) must not exceed the diffuse astrophysical flux measured by the IceCube Collaboration~\cite{Abbasi:2021qfz}.
\item The Galactic neutrino component should be bounded by the flux reported by IceCube’s dedicated analysis of the Galactic plane~\cite{IceCube:2023ame}.
\item The SHDM decay lifetime must be compatible with non-observation of a steady neutrino flux by IceCube~\cite{IceCube:2025ezc} and the Pierre Auger Observatory~\cite{PierreAuger:2019ens} in the energy range relevant for KM3-230213A.
\item The associated gamma-ray flux must not overshoot the integrated upper limits set by gamma-ray and cosmic ray experiments~\cite{PierreAuger:2025jwt,KASCADEGrande:2017vwf,CASA-MIA:1997tns,Fomin:2017ypo,TelescopeArray:2018rbt,Abbasi:2021Z9}.

\end{itemize}

To consistently incorporate these constraints, we define the following likelihood function:
%
\begin{align}\label{eq:lik}
\mathcal{L}(M_\chi, \tau_\chi) =\ 
&\mathcal{P} \left[ \sum_i N_i(M_\chi, \tau_\chi) + N_{\rm back}^{\rm astro}, 1 \right] \nonumber \\
&\times \exp \left[ -\frac{\chi^2_{\gamma} (M_{\chi}, \tau_\chi)}{2} \right] \nonumber \\
&\times \exp \left[ -\frac{f^2_{\rm G} (M_\chi, \tau_\chi)}{2 \sigma_{\rm G}^2} \right] \nonumber \\
&\times \exp \left[ -\frac{\chi^2_{\rm D} (M_\chi, \tau_\chi)}{2 \sigma_{\rm D}^2} \right]
\end{align}
where the sum is over \[ i= \{ \text{KM3NeT/ARCA, IceCube, Pierre Auger} \}~.\] 

The function $\mathcal{P}(\sum_i N_i, 1)$ denotes the Poisson probability for the combined detectors to observe one event given the expected sum $\sum_i N_i(M_\chi, \tau_\chi)$. This approach effectively treats the three experiments as a single, composite detector, with the recorded event in KM3NeT/ARCA representing a detection in one segment of the overall exposure. A similar statistical treatment was adopted in~\cite{KM3NeT:2025ccp} to determine the isotropic flux corresponding to the detection of KM3-230213A. To compute the expected number of events in the energy interval above $10\, \rm PeV$ (denoted as $\Delta E$), we include both extragalactic and Galactic DM contributions. For the extragalactic flux, the expected number of events for each experiment is given by:
\begin{equation}
N^{\rm EG}_{i} = 4\pi T_i \int_{\Delta E} A^{\rm eff}_{i} (E) \frac{d^2 \phi^{\nu}_{\rm EG}}{dE \, d\Omega} (E; M_{\chi}, \tau_\chi) dE \, ,
\end{equation}
%
where $T_i$ and $A^{\rm eff}_i(E)$ denote, respectively, the experimental operational time and the energy–dependent effective area of each experiment. For KM3NeT/ARCA we adopt $T_i = 335\,\mathrm{days}$, while for IceCube we use $T_i = 12.6\,\mathrm{yr}$, with effective areas taken from~\cite{KM3NeT:2025npi,IceCube:2025ezc}. For the Pierre Auger Observatory, we employ the exposure reported in~\cite{PierreAuger:2019ens}. For the Galactic component, we use

\begin{equation}\label{eq:galactic_events}
N^{\rm G}_{i} = T_{i} \int_{\Delta E} A_{i}^{\rm eff}(E) \frac{d\phi^\nu_{\rm G}}{dE d\Omega}(E; M_{\chi}, \tau_\chi, \Omega) F_i(\Omega) dE d\Omega~,
\end{equation}

where $F_i(\Omega)$ is the fractional visibility for each experiment, accounting for a limited field of view~(FoV)~\cite{Arguelles:2022nbl}. The last three terms in Eq.~\eqref{eq:lik} are Gaussian penalty factors, each enforcing the corresponding observational constraint described above. The gamma-ray term, $\chi_{\gamma}^{2} (M_{\chi}, \tau_\chi)$, follows the approach of Ref.~\cite{Chianese:2021jke} and is defined as
\begin{equation}
\chi_{\gamma}^{2}(M_{\chi}, \tau_\chi) = \sum_{j} \left( \frac{\Phi^{\rm G}_{\gamma}(> E_j; M_{\chi}, \tau_\chi)}{\sigma_j} \right)^2,
\end{equation}
where the sum runs over all 95\% CL upper limits from the various gamma-ray experiments, reported in~~\cite{PierreAuger:2025jwt,KASCADEGrande:2017vwf,CASA-MIA:1997tns,Fomin:2017ypo,TelescopeArray:2018rbt,Abbasi:2021Z9}. Each uncertainty is taken as half the reported upper limit, i.e., $\sigma_j = 0.5 \Phi_{95\% \rm UL}$.  Here, $\Phi_{\gamma}(> E_j)$ is the integrated, angle-averaged gamma-ray flux above energy $E_j$, as gamma-ray experiments constrain the energy-integrated flux per solid angle. For the diffuse neutrino flux constraint, we account for the fact that IceCube has not observed any events in the $5-100$~PeV energy range and has set a 90\% CL upper limit of $\Phi^{\rm IC}_{90\%} \le 7.79 \times 10^{-16} \rm cm^{-2} s^{-1} sr^{-1}$~\cite{Abbasi:2021qfz}. 
The corresponding penalty term is
\begin{equation}
\chi^2_{D} = \left(\frac{\Phi_{\nu}^{\rm G + EG}(> E_\nu; M_\chi, \tau_\chi)}{\sigma_D}\right)^2,
\end{equation}
where $\sigma_D = 0.5 \Phi^{\rm IC}_{90\%}$, and the total expected neutrino flux is obtained by integrating between 5 PeV and 100 PeV the sum of the Galactic~(Eq.~\ref{eq:total_flux_galactic}) and Extra-Galactic~(Eq.~\ref{eq:total_flux_extragalactic}) contribution to the angle-averaged diffuse flux. IceCube has also provided a measurement of the Galactic neutrino flux using a template likelihood analysis of cascade (shower) events in the energy range $\sim 1-10^3$~TeV~\cite{IceCube:2023ame}. The measured neutrino intensity at $E = 1,\rm PeV$ is $E^2 \phi^{\rm meas}_{\nu+\bar \nu}(E) \sim 3.89 \times 10^{-9} \rm GeV cm^{-2} s^{-1}$, integrated over the sky, with a relative  statistical uncertainty  reported of $\sigma_G \simeq 30$\%~\cite{IceCube:2023ame}.  
To quantify the corresponding Gaussian penalty, we define the Galactic fractional DM contribution to the measured flux at 1 PeV
and require that this fraction~$(f_G)$ be consistent with zero within the stated uncertainty. 

$N_{\rm back}^{\rm astro}$ represents the contribution from a potential astrophysical background to the total number of neutrino events observable by the detectors. This contribution is both theoretically and observationally well-motivated, though it is subject to substantial uncertainties that could significantly influence our analysis~\cite{Ehlert:2023btz,Aloisio:2015ega,KM3NeT:2025npi,Condorelli:2022vfa,Rossoni:2024ial,Winter:2022fpf,AlvesBatista:2018zui,Heinze:2019jou,KM3NeT:2025lly}. In order to account for it while remaining both conservative and model–independent, we adopt $N_{\rm back}^{\rm astro} = 1$. This choice corresponds to attributing the KM3-230213A  event to the astrophysical background.  This assumption is conservative for several reasons: higher values of $N_{\rm back}^{\rm astro}$ would result in even stronger constraints, and astrophysical backgrounds—such as cosmogenic or in-source neutrinos—could also contribute to the Gaussian penalty terms in the likelihood, further tightening the limits. 
Future analyses might be able to rely on realistic astrophysical predictions, provided that observational evidence for an astrophysical flux at these energies becomes available.
We define a Test Statistic~$(\rm TS)$ as 
\begin{equation}
\mathrm{TS}(M_\chi, \tau_\chi) = 2 \ln \left( \frac{\mathcal{L}(\bar \tau_\chi \longrightarrow +\infty)}{\mathcal{L}(M_\chi, \tau_\chi)} \right)~,
\end{equation}
and derive a one-sided 95\% confidence level (CL) lower limit on the dark matter lifetime as a function of its mass by imposing ${\rm TS} = 2.71$  \cite{LHAASO:2022yxw,Cowan:2010js}.

\section{Results}
\label{sec:results_1}

Figure~\ref{fig:constraints_DM_lifetime} shows the resulting lower limits on $\tau_\chi$ as a function of $M_\chi$. The left and right panels refer to $\chi \rightarrow \nu \bar{\nu}$ and $\chi \rightarrow b \bar{b}$ channels, respectively. 
In general, the decay channel $\chi \rightarrow b\bar{b}$ is more strongly constrained due to its higher gamma-ray yield.
Remarkably, even under our conservative assumptions, the KM3-230213A event allows us to set the most stringent constraints to date on $\tau_\chi$ for both decay channels considered. Specifically, in the case of $\chi \longrightarrow b \bar{b}$, we find an improvement of up to a factor of $\sim 1.5$ for $M_\chi \lesssim 10^{11}, \rm GeV$ compared to Ref.~\cite{Das:2023wtk}, while our results are comparable at higher masses. Furthermore, for $M_\chi \lesssim 10^{9}\,  \rm GeV$, our constraints surpass those obtained by the LHAASO collaboration~\cite{LHAASO:2022yxw}.
Our findings reinforce previous results~\cite{PierreAuger:2022ubv,PierreAuger:2022jyk,Berat:2022iea,Das:2023wtk}, underscoring the power of gamma-ray observations in constraining a broad class of beyond-the-Standard-Model scenarios. Further improvements in gamma-ray upper limits at the highest energies~\cite{Castellina:2019irv}  will provide complementary information and play a decisive role in probing or excluding SHDM-induced signatures even further. 
In the case of $\chi \rightarrow \nu \bar{\nu}$, the limits improve by up to a factor of $\sim 3$ compared to Ref.~\cite{Chianese:2021jke}, and by an even larger factor relative to those reported in Ref.~\cite{Arguelles:2022nbl}. This enhanced sensitivity arises from a higher neutrino yield associated with this decay channel. The galactic neutrino flux, in the energy range $\gtrsim 10\, \rm PeV$, is crucial to falsify the SHDM decay hypothesis even further. Therefore, future experiments such as GRAND, RNG-0, IceCube Gen 2 and POEMMA promise unprecedented sensitivity, potentially enabling the extension of these constraints to higher energies~\cite{GRAND:2018iaj,IceCube-Gen2:2020qha,POEMMA:2020ykm,2025JInst..20P4015A}.

\section{Conclusions}
\label{sec:conclusions}


In this manuscript, we have utilized the highest-energy neutrino ever detected, KM3-230213A, to establish the most stringent constraints on the SHDM decay scenario to date. Even adopting a conservative approach, we derive robust lower bounds on the dark matter lifetime, $\tau_{\chi} \gtrsim 5\cdot 10^{29}-10^{30} \rm s$ for $10^{7}\, \rm GeV \lesssim m_{\chi} \lesssim 10^{17}\, \rm GeV$, slightly varying based on the decay channel considered. Future analyses employing detailed spatial distribution templates are expected to further refine these constraints, thereby placing neutrino astronomy at the forefront of probing SHDM models.

We anticipate that imminent advancements in neutrino observatories -- most notably KM3NeT/ARCA and IceCube-Gen2 -- with their enhanced exposure and sensitivity toward the Galactic neutrino background, complemented with upcoming ultra-high-energy gamma-ray observations, such as those from AugerPrime~\cite{Castellina:2019irv}, will jointly play a crucial role in refining constraints on the SHDM decay hypothesis, potentially providing decisive evidence against hadronic decay channels or tightening constraints on leptonic channels.

\begin{acknowledgments}
\textit{Acknowledgments}: AA acknowledges the support of the project ``NUSES - A pathfinder for studying astrophysical neutrinos and electromagnetic signals of seismic origin from space'' (Cod.~id.~Ugov: NUSES; CUP: D12F19000040003). The work of RA and CE has been partially funded by the European Union – Next Generation EU, through PRIN-MUR 2022TJW4EJ and by the European Union – NextGenerationEU under the MUR National Innovation Ecosystem grant ECS00000041 – VITALITY/ASTRA – CUP D13C21000430001. 
\end{acknowledgments}

\bibliography{biblio}
\bibliographystyle{apsrev4-2}

\end{document}

%% file: preamble.tex
\usepackage{aas_macros}
\usepackage{amsmath,amssymb,amsfonts}
\usepackage[english]{babel}
\usepackage{gensymb}
\usepackage{graphicx}
\usepackage{orcidlink}
\usepackage{physics}
\usepackage{soul}
\usepackage[dvipsnames]{xcolor}

\usepackage{hyperref}
\hypersetup{
    colorlinks=true,
    linkcolor=blue,
    citecolor=blue,
    filecolor=magenta,      
    urlcolor=blue,
    pdftitle={Aloisio+,2025},
    pdfpagemode=FullScreen,
    }